# On the recent large EQ of Karpathos, Greece (April 1st, 2011, Ms = 6.7R) and its similarities to the large EQ of the January 1st, 2006 (Ms = 6.9R) at East of Kythira, Greece.

Thanassoulas[1], C., Klentos[2], V., Verveniotis, G.[3], Zymaris, N.[4]

1. Retired from the Institute for Geology and Mineral Exploration (IGME), Geophysical Department, Athens, Greece.
   e-mail: thandin@otenet.gr - URL: www.earthquakeprediction.gr

2. Athens Water Supply & Sewerage Company (EYDAP),
   e-mail: klenvas@mycosmos.gr - URL: www.earthquakeprediction.gr

3. Ass. Director, Physics Teacher at 2nd Senior High School of Pyrgos, Greece.
   e-mail: gver36@otenet.gr - URL: www.earthquakeprediction.gr

4. Retired, Electronic Engineer.

**Abstract**
The recent large EQ that occurred in Greece (April 1st, 2011, Ms = 6.7R) is investigated as far as it concerns regional geophysical features of the Greek territory. In particular, the EQ location is compared to the deep lithospheric fracture zones and faults, derived from the analysis of the corresponding earth's gravity field; to the current seismic potential map determined from the study of the past seismicity, while its time of occurrence is compared to the peaks of the M1 and diurnal tidal waves. The detailed investigation of the earth's electric field that was recorded by the ATH monitoring site, located in Athens, revealed the presence of short train like pulses type electric seismic precursory signals which were generated short (1 – 2 days) before the EQ occurrence time. Moreover, the analysis (for T = 1 day) of the earth's oscillating electric field, that was simultaneously recorded at PYR and ATH monitoring sites, revealed that the "strange attractor like" seismic electric precursor preceded for 1 – 2 days the EQ occurrence time. A similar behavior was observed from the large EQ of the East Kythira (Ms = 6.9R, January, 8th, 2006). It is concluded that the same regional tectonic mechanism controlled and generated both the analyzed EQs while the adopted geophysical earth models have been validated.

## 1. Introduction.

On 1st of April, 2011 a large EQ of Ms = 6.7R (NOA) occurred at the South-Eastern part of the Greek territory, between the Creta and Karpathos island. Those kind of large EQs, in Greece, mostly take place at the Southern Hellenic Arc, where the African plate, due to its North – East motion, subducts the South – Western drifting Aegean plate. A generalized plate motion model, after McKenzie (1972, 1978) for the regional area, is presented in figure (1).

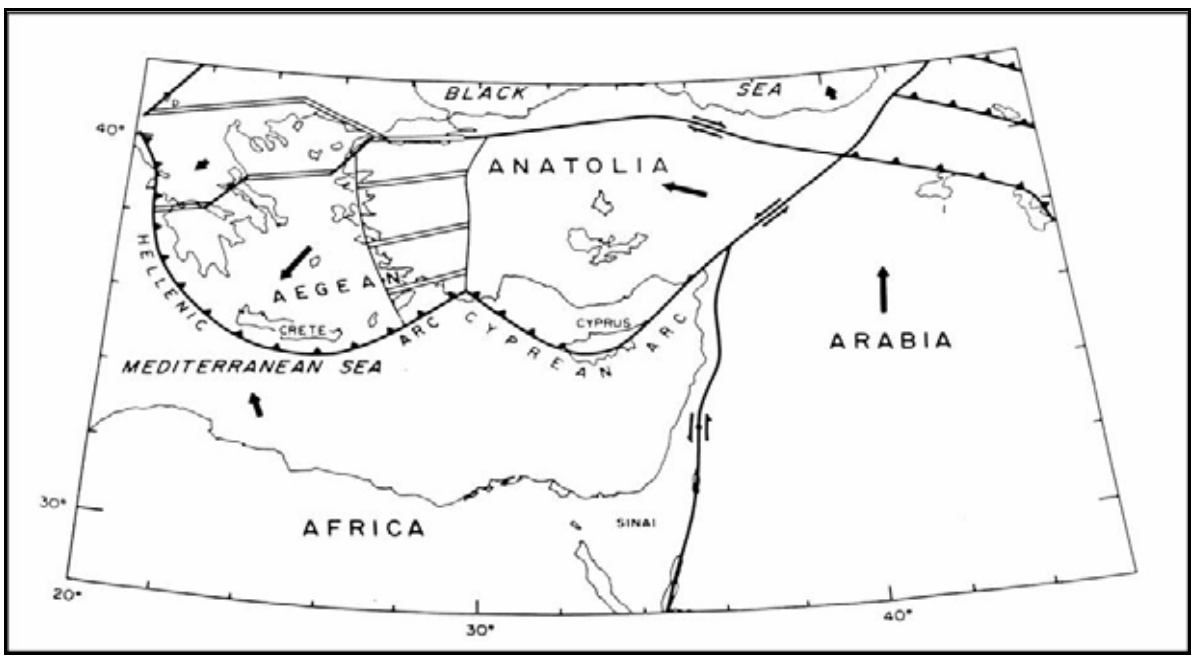

Fig. 1. McKenzie's (1972, 1978) lithospheric plate model proposed for the Greek territory.

The mentioned afore EQ took place between the eastern part of Creta and the Karpathos Island (see fig. 2).

## 2. Data presentation

The exact location parameters of the particular EQ are presented by EMSC in the following figure (2) along with the tectonic plate boundaries. It must be noted that the EMSC determines magnitudes in Mw while the NOA in Ms.

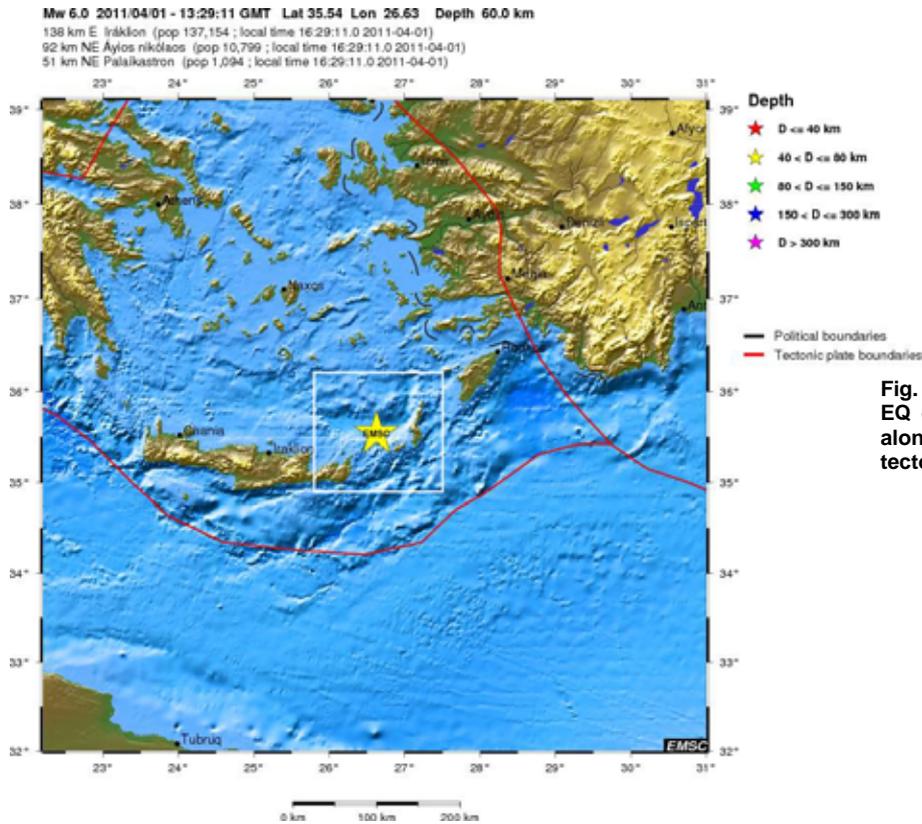

Fig. 2. Location (yellow star) of the large EQ of the 1$^{st}$ of April 2011 of Ms = 6.7R along with the boundaries (red line) of the tectonic plates (after EMSC).

More geophysical details about the location of the particular EQ can be obtained if it is compared to the deep lithospheric fracture zones / faults that characterize the regional seismogenic area. These zones / faults were mapped by converting the earth's gravity field into its horizontal gradient (Thanassoulas, 1998, 2007) and by compiling the corresponding maps. Actually, a large EQ will occur at a place where a large faulting zone exists, which will be activated by the presence of excess tectonic stress load. Consequently, the epicenter of the future EQ will coincide (in geological scale) to the location of the activated fault zone. This is demonstrated, for the particular EQ, in the following figure (3) left. The lithospheric fracture zones of the Greek territory are presented by brown lines while the EQ epicenter is shown blue concentric circles.

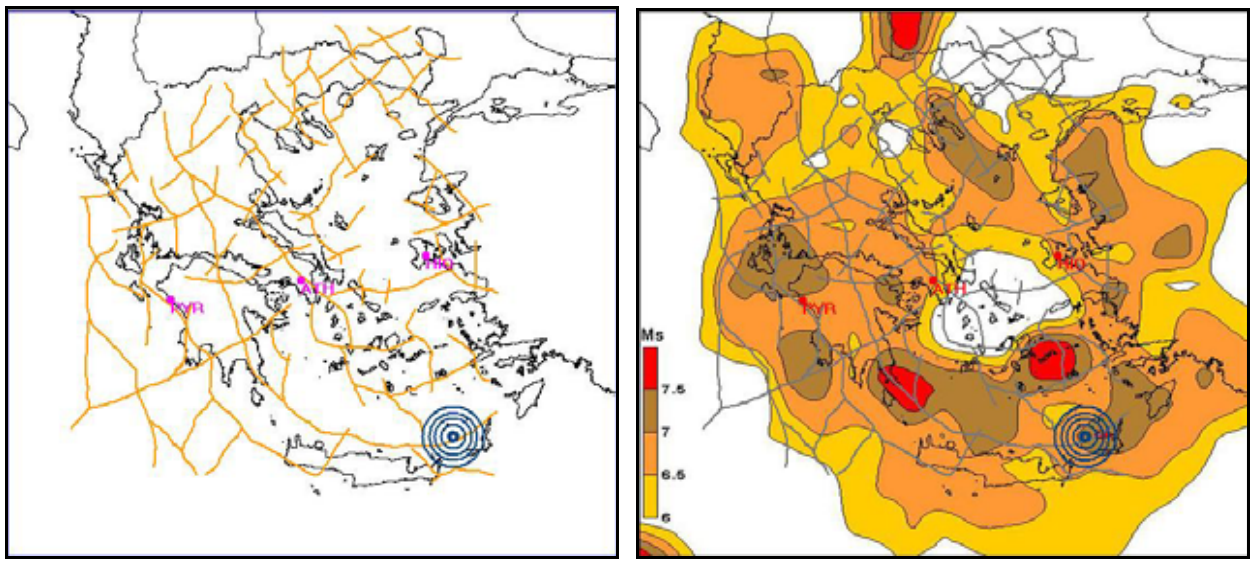

**Fig. 3. Lithospheric fracture zones / faults (brown lines left) and seismic potential map (right) calculated for 2010. Blue concentric circles denote the EQ location.**

Moreover, the occurrence of a large EQ means that a large amount of strain energy release takes place. Therefore, that energy must have been accumulated in the regional seismogenic area during the past time. Maps of stored strain energy can be compiled by analyzing the past seismic history of any seismogenic area by using the "lithospheric seismic energy flow model" (Thanassoulas 2008, 2008a; Thanassoulas et al. 2001, 2003, 2008). Thus, large EQs are expected to occur at places where large amount of strain energy has been identified. In the present case, the particular EQ location is compared (fig. 3, right) to the seismic potential map of Greece compiled for the year 2010 (Thanassoulas and Klentos, 2010). The specific EQ occurred at the boundaries between Ms = 6.5R and Ms = 7.0R.

Although the preparation of a large EQ takes some decades of years before its occurrence, during its final phase and short before its occurrence drastic mechanical changes take place in the rock formation present in its focal area. Micro-cracking accelerates, EQ nucleation is initiated and at a certain level of cracking size, electric signals are generated (Thanassoulas, 2007, 2008b). In this case, train-like pulses characteristic electric preseismic signals were generated a few hours in the same day before the EQ occurrence while the very same type electric signals had been presented 24 hours before, in the previous day. In the following figure (4) a two days period (20110331 – 20110401, yyyymmdd) recording is presented for the earth's electric field for its N-S and E-W components. The red circles indicate the presence of the preseismic electric signals while the red bar indicates the EQ occurrence time.

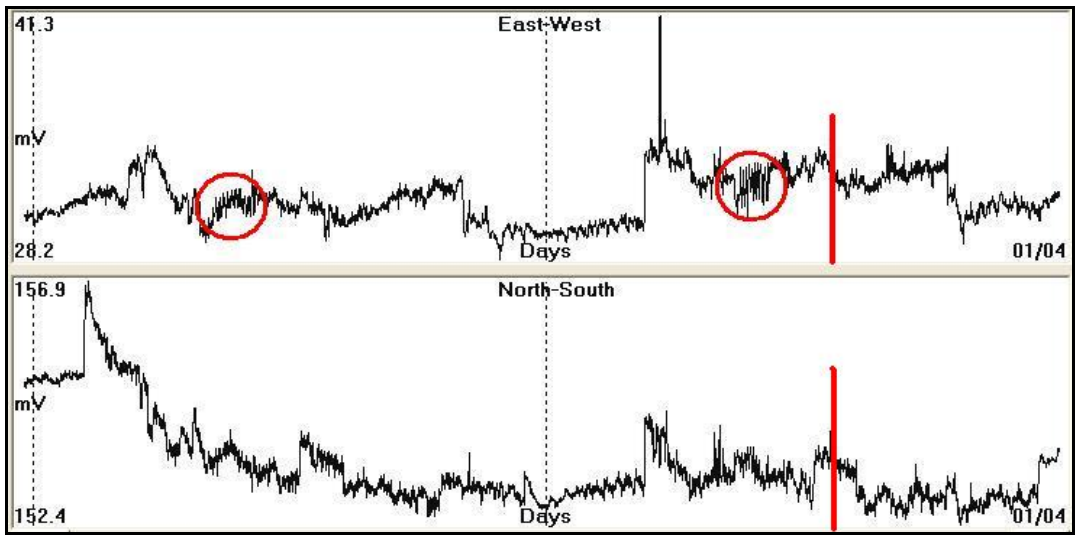

**Fig. 4.** Earth's electric field recorded, by ATH monitoring site in Athens, for the time period of 20110331 – 20110401 (yyyymmdd). Red circles denote the presence of train like type electric pulses preseismic signals. Red bar denotes the large EQ occurrence time.

A better enlarged presentation of the preseismic electric signals is given in the following figures (5, 6). Figure (5) shows the recorded preseismic electric signal one day before the EQ occurrence, while the figure (6) shows the recorded preseismic electric signal a few hours before the EQ. The horizontal red bar indicates the duration of the signal.

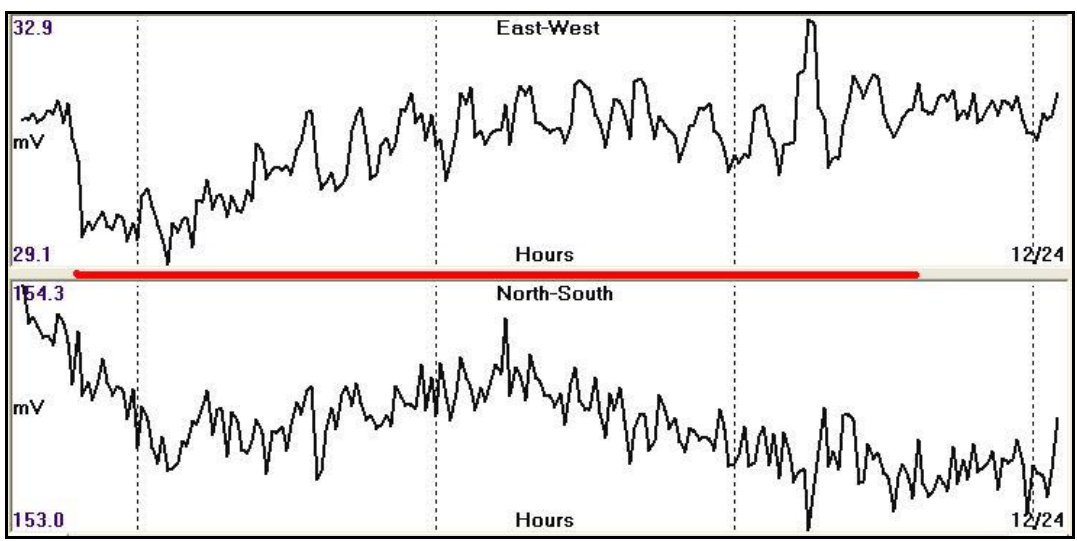

**Fig. 5.** Zoom-in of the left red circle of figure (4). The red bar indicates the signal duration of the 20110331.

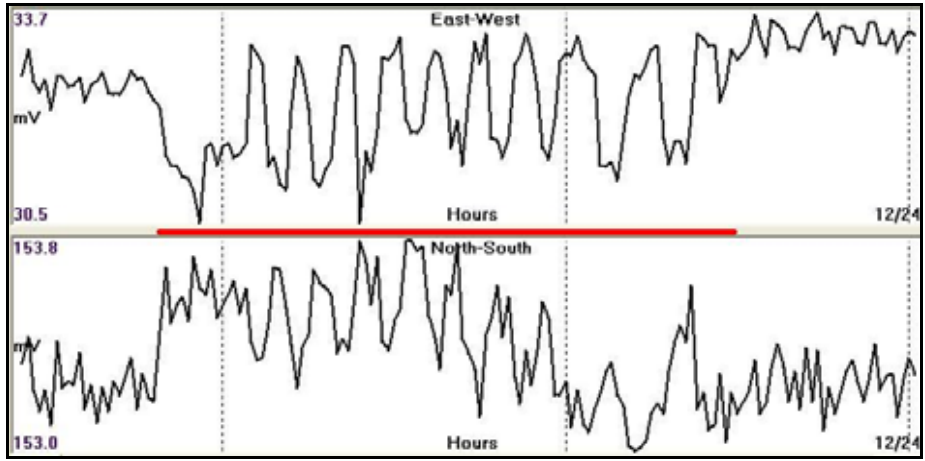

**Fig. 6. Zoom-in of the right red circle of figure (4). The red bar indicates the signal duration of the 20110401**

An interesting feature of the presented electric field in figure (4) is the time of occurrence of the preseismic signal. Both signals were initiated about the same time. Therefore, the triggering mechanism for the generation of these electric signals must present a periodicity of T = 24 hours. The latter leads to adopt the tidally triggered lithospheric oscillation mechanism, due to the diurnal tidal variation. At the amplitude maxima of its oscillation, the lithosphere reaches maxima of stress load and therefore, if it is at critical stress load conditions, just before the EQ occurrence, temporary fracturing takes place during its oscillation peaks and as a result of it electric signals are emitted during these peaks. The specific electric signals generating mechanism is demonstrated in the following figures (7, 8) separately for each corresponding day of recording. The preseismic electric field recording (black line) has been presented along with the corresponding diurnal tidal (red line) variation.

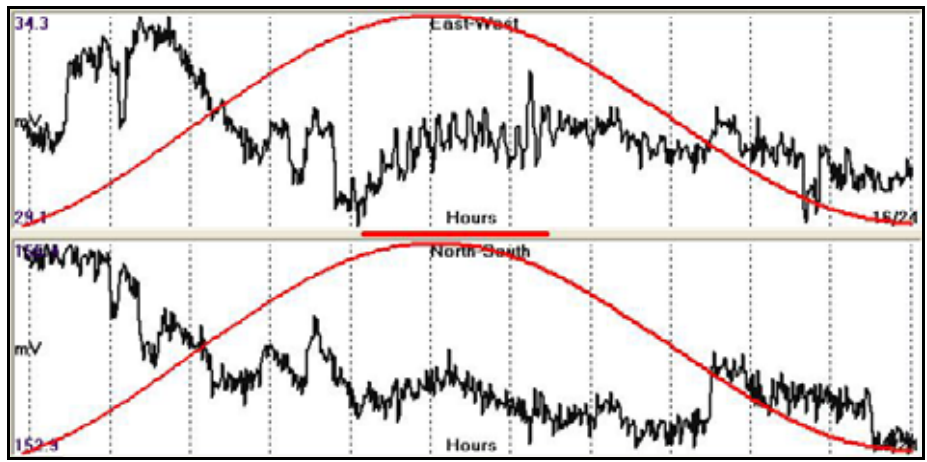

**Fig. 7. Recorded electric signal (black line) on the 20110331 is shown along with the corresponding diurnal tidal variation (red line). The red horizontal bar indicates the electric signal duration.**

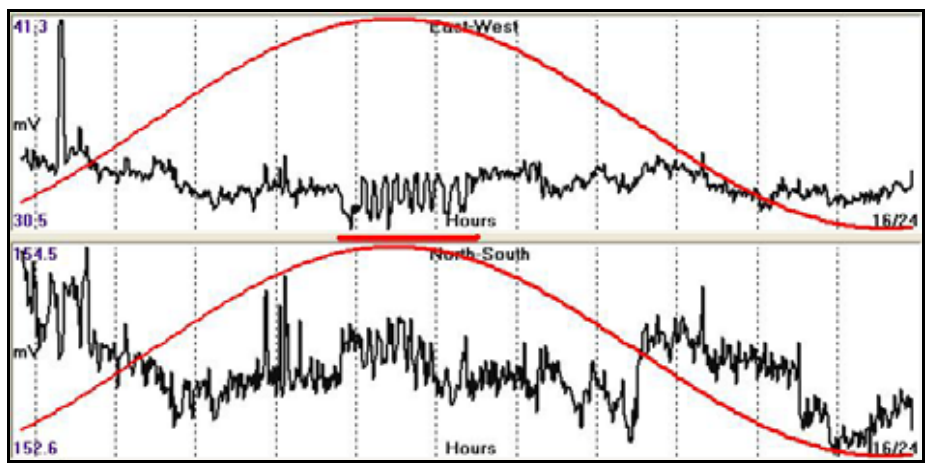

**Fig. 8. Recorded electric signal (black line) on the 20110401 is shown along with the corresponding diurnal tidal variation (red line). The red horizontal bar indicates the electric signal duration.**

**The coincidence of the preseismic electric signals generation time to the same day diurnal tidal variation amplitude peak is very clear. Moreover it must be noticed that although the dominant component of the diurnal variation is in general the K1 (T = 24 hours) component, in the present case the K2 (T = 12 hours) prevails.**

**Another interesting seismic electric precursor that preceded the Karpathos large EQ is the "strange attractor like" one. When the focal area is at a very critical dynamic state before fracturing, as long as the lithosphere oscillates due to tidal triggering, different types of electric signals of a wide variety of amplitudes are generated. Of specific interest is the oscillating earth's electric field component with T = 1 day. It has been shown (Thanassoulas, 2007, Thanassoulas et al. 2008a, 2009, 2009a) that the specific oscillating electric field, short before the EQ occurrence, generates ellipses, as time functions, by the intersection of the electric field intensity vectors observed at two different monitoring sites. If the future focal area is not at critical stress-charge state then random hyperbolas are generated. In the present case the two monitoring sites are located in Athens (ATH) and Pyrgos (PYR) towns of Greece (see fig. 3 left and following maps).**

**Phase maps have been compiled from the data obtained (6) days before the EQ occurrence and (2) days after (including the occurrence day). The aim of this work is to investigate the presence of the "Strange attractor like" seismic electric precursor before the EQ occurrence time. In a positive case it has a predictive value towards the utilization of a short-term earthquake prediction. The following maps presented in figures (9, 10, 11, 12) have been compiled for T = 1 day of the Earth's oscillating electric field.**

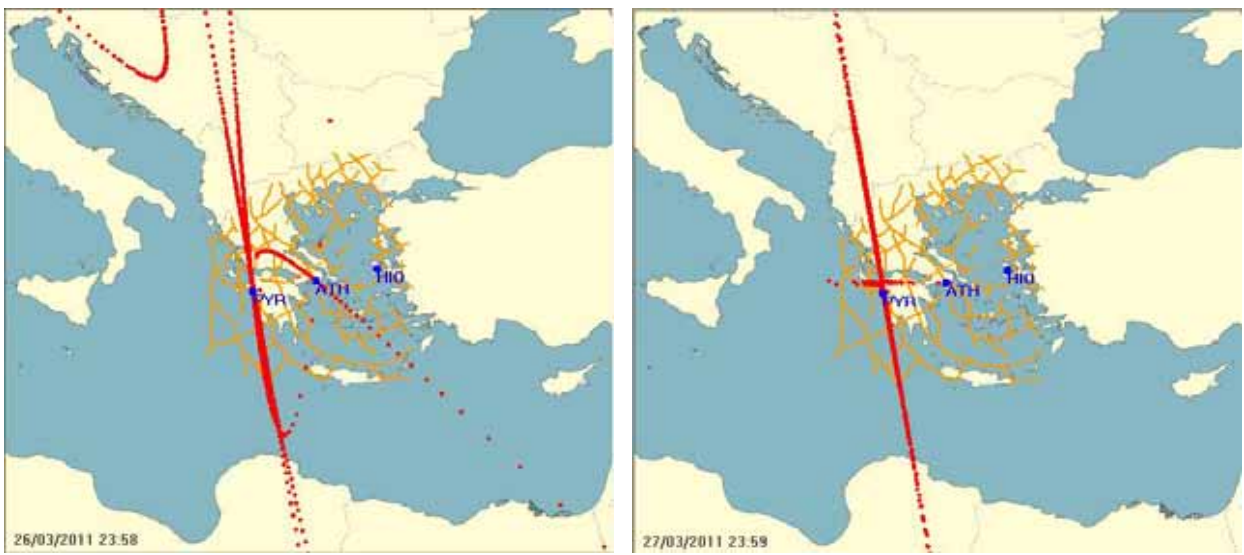

**Fig. 9. Phase maps determined for 20110326 (left) and 20110327 (right).**

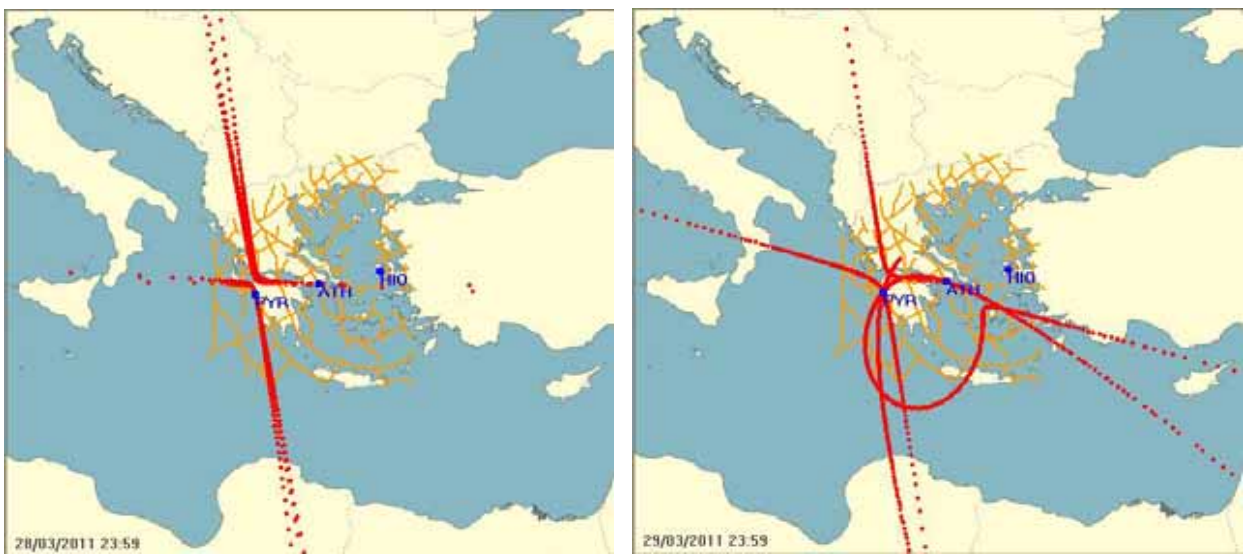

**Fig. 10. Phase maps determined for 20110328 (left) and 20110329 (right).**

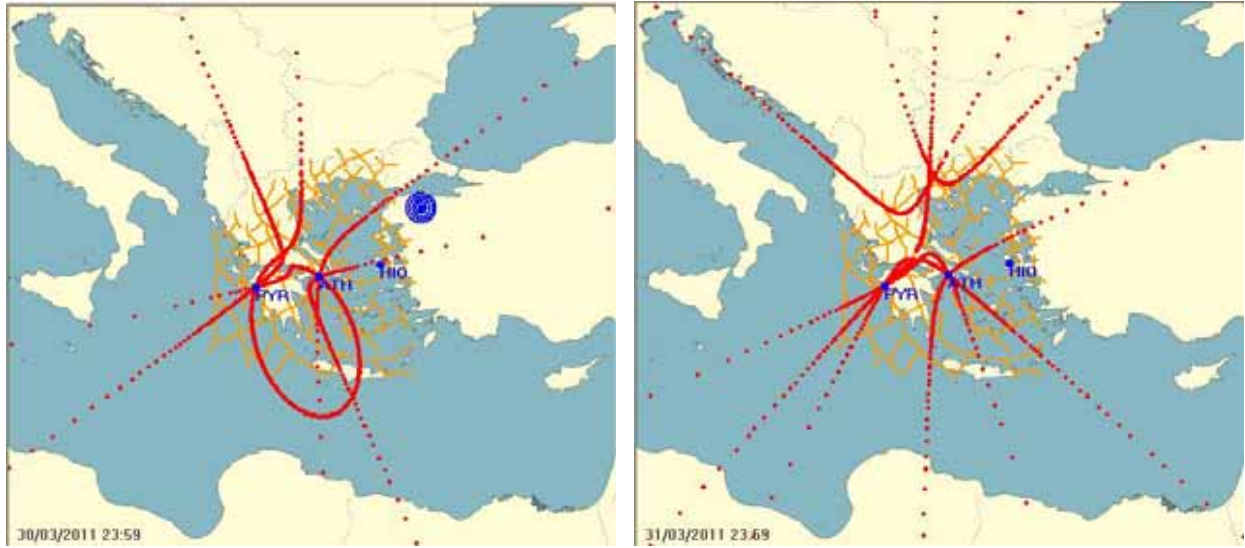

**Fig. 11. Phase maps determined for 20110330 (left) and 20110331 (right).**

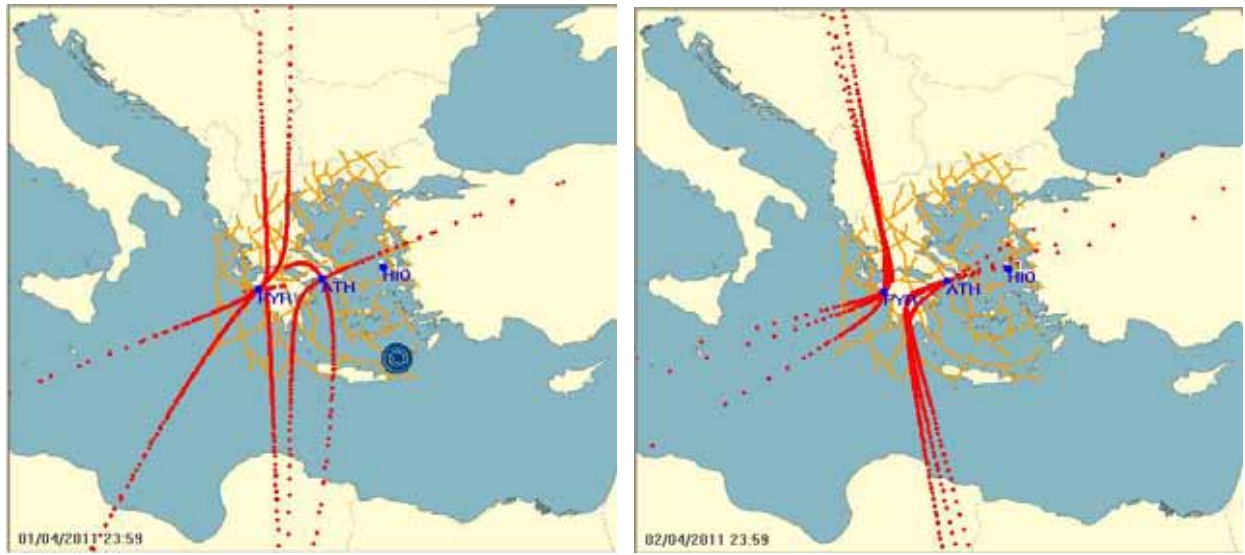

**Fig. 12. Phase maps determined for 20110401 (left) and 20110402 (right).**

The "strange attractor like" seismic electric precursor (presence of ellipses) appears on March 29th and March 30th. A small EQ (blue concentric circles) with magnitude of Ms = 4.5R occurs on the March 30th while the ellipses are still present. On March 31st the ellipses vanish and the next day, April 1st, 2011 the Karpathos large EQ (blue concentric circles) takes place. On April 2nd 2011 the earth's electric field returns back to normal, generating typical hyperbolas.

It is worth to notice that the EQ did occur two (2) days after the vanishing of the ellipses, which comply with similar results found from other earlier presented cases (Thanassoulas et al. 2009).

Finally the Karpathos large EQ time of occurrence is compared to the time when the M1 and diurnal tidal waves achieve their maximum amplitude. Similar studies (Thanassoulas, 2007) has shown that "the by chance" $P_{ch}$ coincidence of the time of occurrence of a large EQ (Ms>5.5R) with any day of the half period (7 days) of M1 is $P_{ch}$ = 14.8% while a deviation mean value was determined as 1.18 days from the time of occurrence of the M1 maximum amplitude, after the analysis of 40 large EQs for the period 1995 – 2001. The latter analysis indicated a 39.47% for a zero (0) value for the deviation. The Karpathos EQ deviates from the M1 tidal peak amplitude time for **1.5 days**. A comparison of the Karpathos EQ occurrence time (red bar) to the M1 tidal component is shown in the following figure (13).

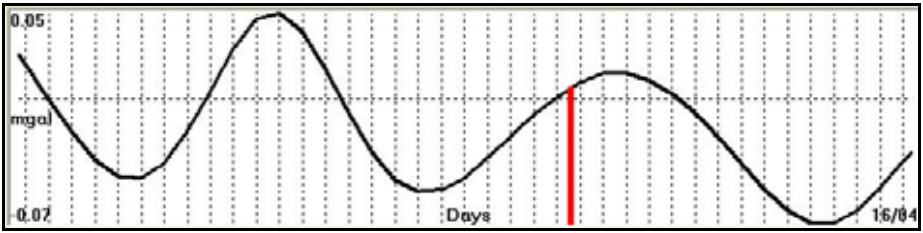

Fig. 13. Time of occurrence of Karpathos EQ (red line) is compared to the M1 (black line) tidal component. Deviation is at the range of 1.5 days.

A similar analysis has been made (Thanassoulas, 2007) for the diurnal tidal variation. The "by chance" Pch probability for a large earthquake (Ms>5.5R) to occur on a diurnal tidal peak is Pch = 16.1% (+/- 1 hour) while the analysis of 70 earthquakes for the period 1964 to 2001 indicated that the mean deviation value from the tidal peak of the K1 equals to 92.6 minutes, that is almost 1.5 hours, while by adopting a deviation window of one hour the 37.1% of the earthquakes fall in this window. The comparison of the Karpathos large EQ occurrence time (red bar) to the diurnal tidal is shown in the following figure (14).

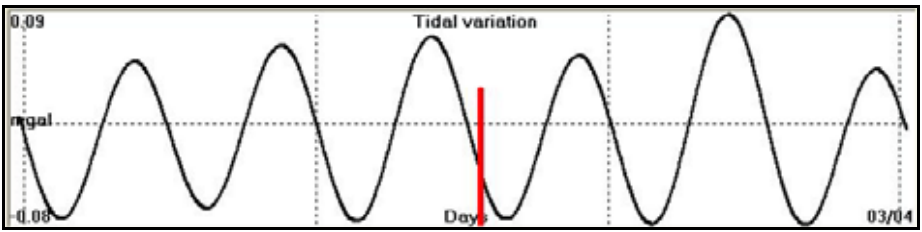

Fig. 14. Time of occurrence of Karpathos EQ (red line) is compared to the diurnal (black line) tidal component. Deviation is at the range of 2 hours.

### 3. Discussion - Conclusions.

The presented Karpathos EQ case has justified earlier studies concerning the lithospheric fracturing and oscillation, the stress load charge determination of seismogenic areas, the generation of preseismic electric signals and the presence of the "strange attractor like" preseismic electric precursor, as well as the tidal triggering mechanism of the earthquakes. All these could be argued on the basis that it is only a single case and therefore it represents no statistical significance.

The problem with all studies and analyses of large EQs (Mw>6.5R) in a certain seismogenic area, as Greece is, is that the number of large EQs for a certain period is very small and therefore, intrinsically, it is prohibited to apply robust statistical methods based on large numbers of samples. Thus, either the term "large" is lowered to smaller magnitudes so that the statistical samples number increases or another similar in behavior example of large EQ is looked for in order to foster the results of the initial study.

In the present case, a large (Ms = 6.9R) earthquake (East Kythira, 20060108) had occurred that presents a very similar behavior as the one of the Karpathos EQ. Following is a comparative presentation of both EQs. In figure (15) left, it is shown that both EQs, of East kythira (left) and Karpathos (right), did occur at the same lithospheric fracture zone. In figure (15) right, the East Kythira EQ coincides in location with an area of seismic potential which is capable of generating EQs larger than 7.5R while the Karpathos EQ coincides to an area capable of generating EQs larger than 7.0R.

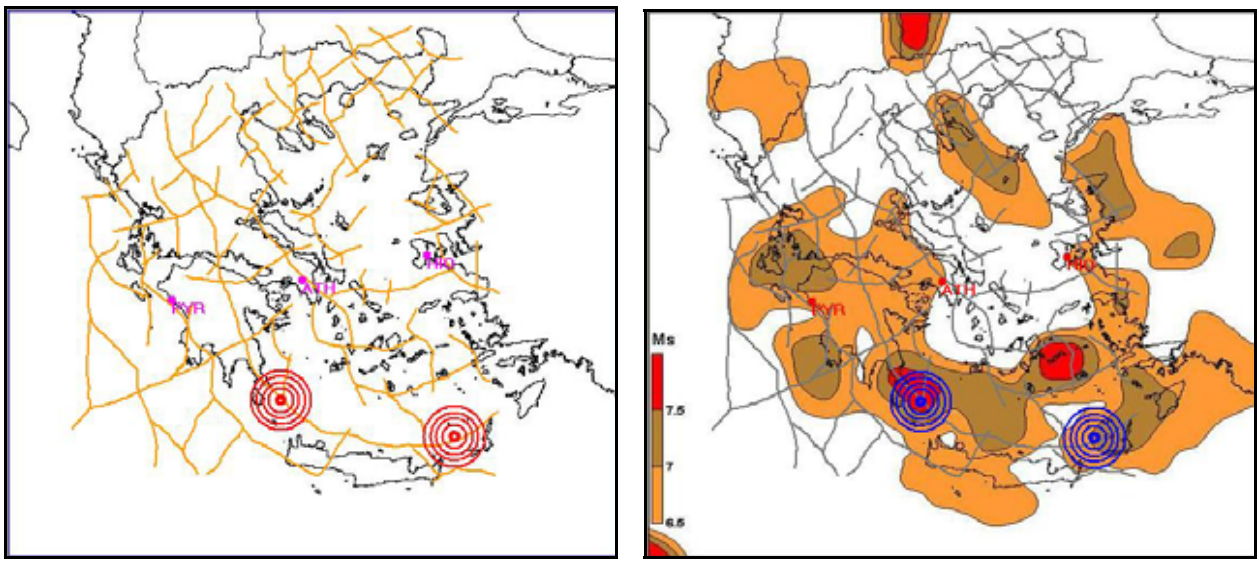

**Fig. 15. Comparative presentation of the East Kythira (left EQ) and Karpathos (right EQ) vs. lithospheric fracture zones / faults (brown lines left) and seismic potential map (right) calculated for 2010. Red and blue concentric circles denote the EQs location.**

Both maps were validated by the occurrence of the two large EQs.

Preseismic electric signals of train like type pulses were generated by both earthquakes. In the following figure (16) the recorded signals generated by the East Kythira EQ are presented. The pulses started almost 1.5 hours before the EQ and ended almost 2 hours after.

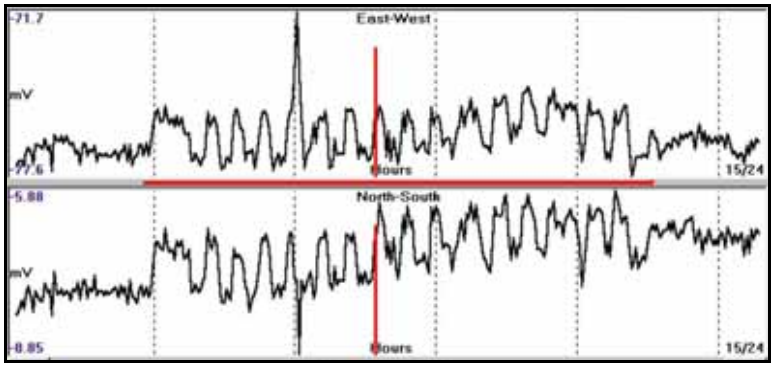

Fig. 16. Preseimic electric signal (train like type pulses) initiated before the East Kythira EQ of 20060108 (Ms = 6.9R).

The Karpathos EQ did generated the very same type of signals for two consecutive days (one day before and during the EQ occurrence day). The Karpathos preseismic electric signal is shown in the following figure (17).

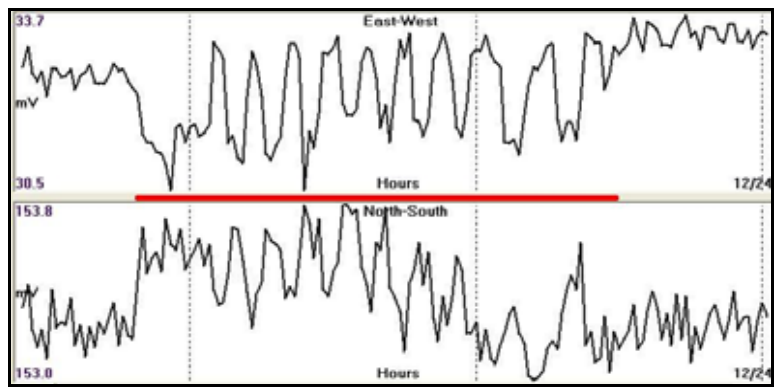

Fig. 17. Preseimic electric signal (train like type pulses) initiated before the Karpathos EQ of 20110401

From figures (16, 17) the pulse length was calculated as:

East Kythira, 20060108:    pulse length = 12.085 minutes

Karpathos 20110401:    pulse length = 8.286 minutes

The observed difference in the pulse length could be attributed to: the different observed magnitudes, to the different physical activated mechanisms in the focal area, to different geological and tectonic conditions and to the different stress and strain loads met in the focal areas.

Another interesting observation is the fact that both EQs generated seismic precursory electric signals that coincide with the same day tidal diurnal variation peak. That comparison is shown in the following figure (18). It is evident that the lithosphere generates electric signals each time, being at critical stress load conditions, it reaches a peak of oscillating strain, which, in this case, is due to the diurnal tidal oscillation.

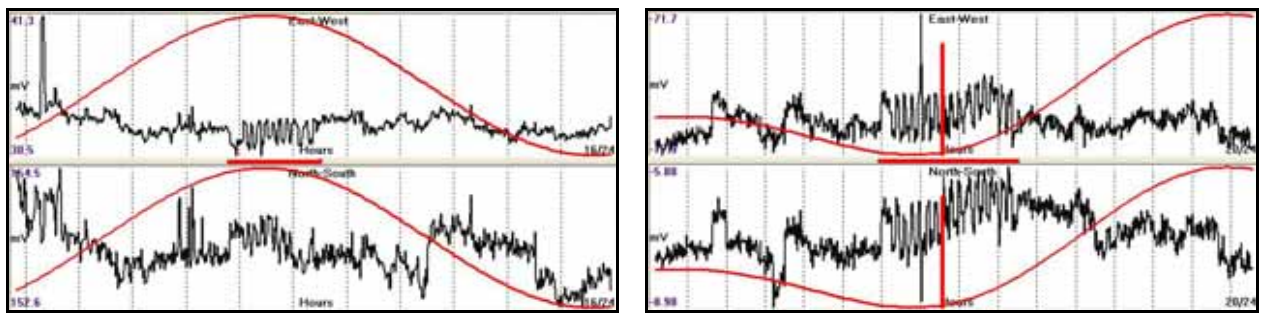

Fig. 18. Karpathos EQ on 20110401 (left) and East Kythira EQ on 20060108 (right) SES compared to the diurnal tidal wave.

Furthermore, the compiled phase maps of the recorded earth's electric field at PYR and ATH monitoring sites behaved in the same way as in the case of Karpathos EQ. The compiled maps for the case of East Kythira EQ are shown in the following figures (19, 20, 21, 22, 23).

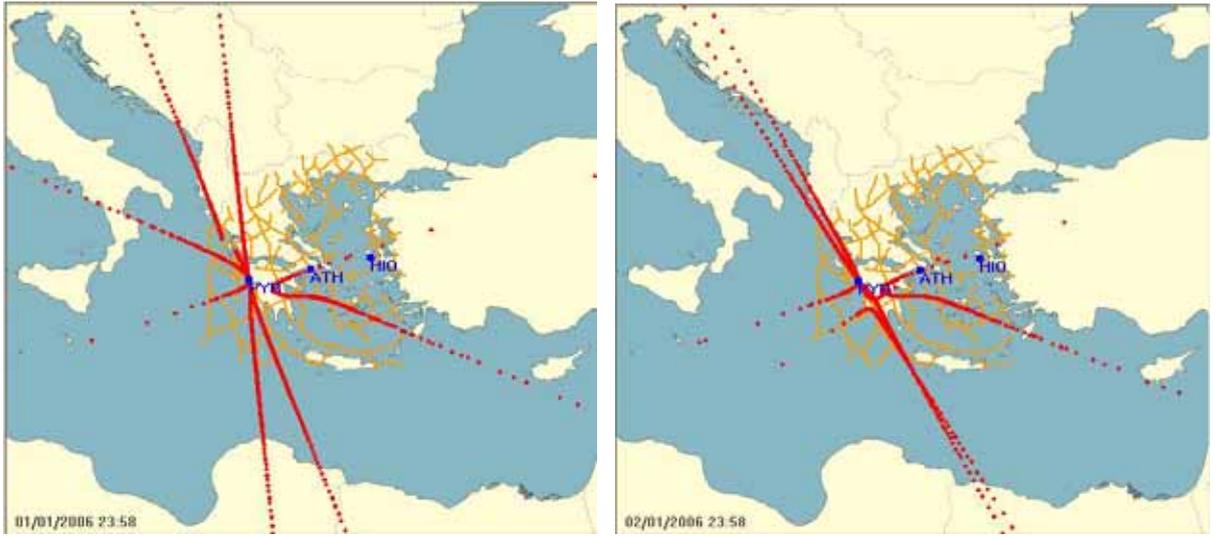

**Fig. 19. Phase maps determined for 20060101 (left) and 20060102 (right).**

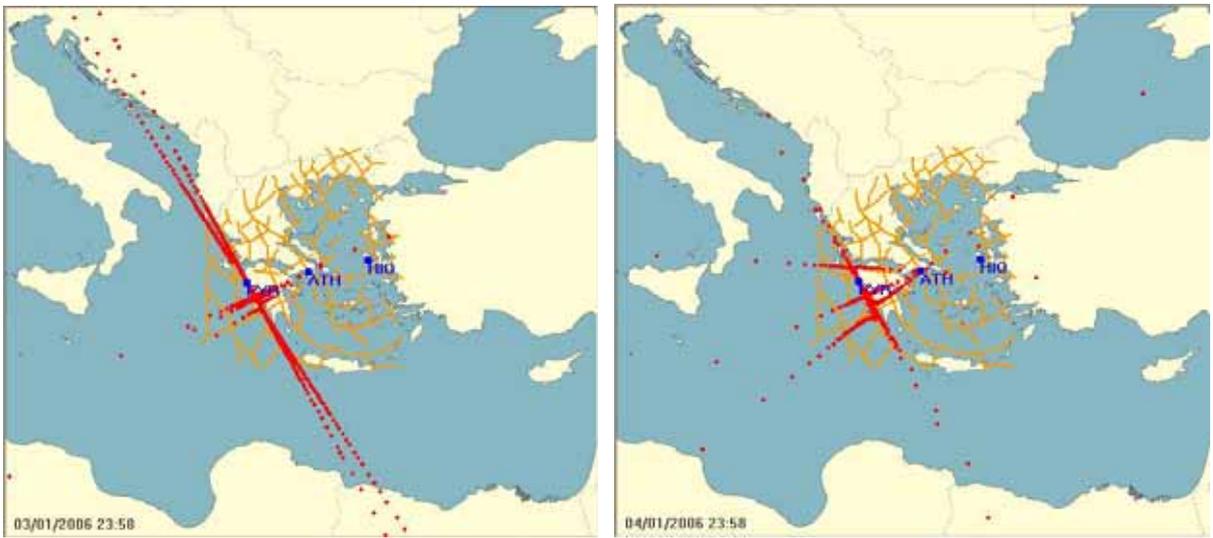

**Fig. 20. Phase maps determined for 20060103 (left) and 20060104 (right).**

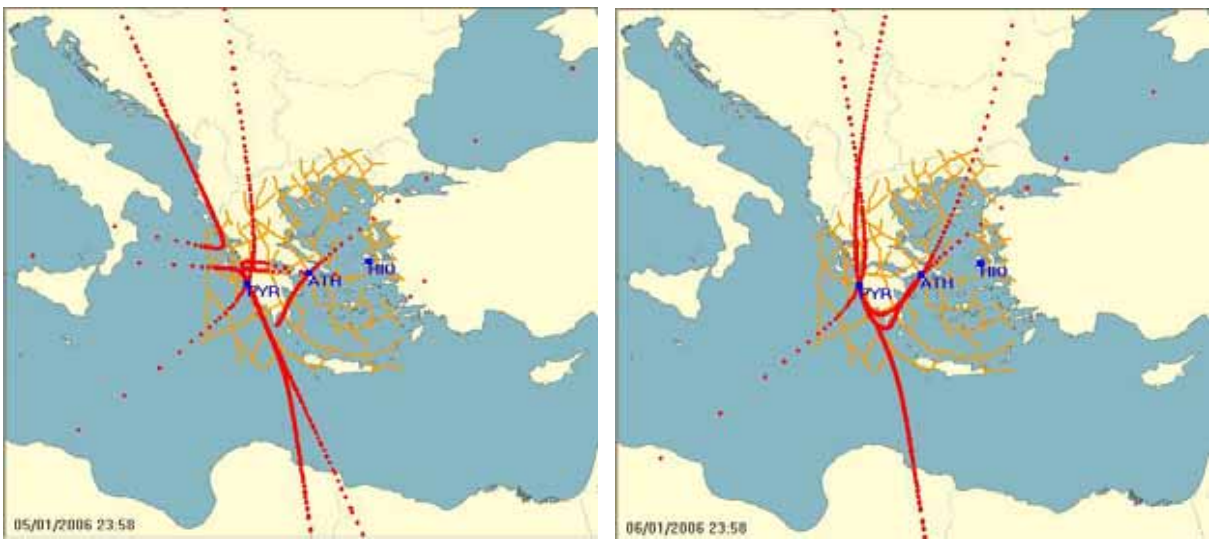

**Fig. 21. Phase maps determined for 20060105 (left) and 20060106 (right).**

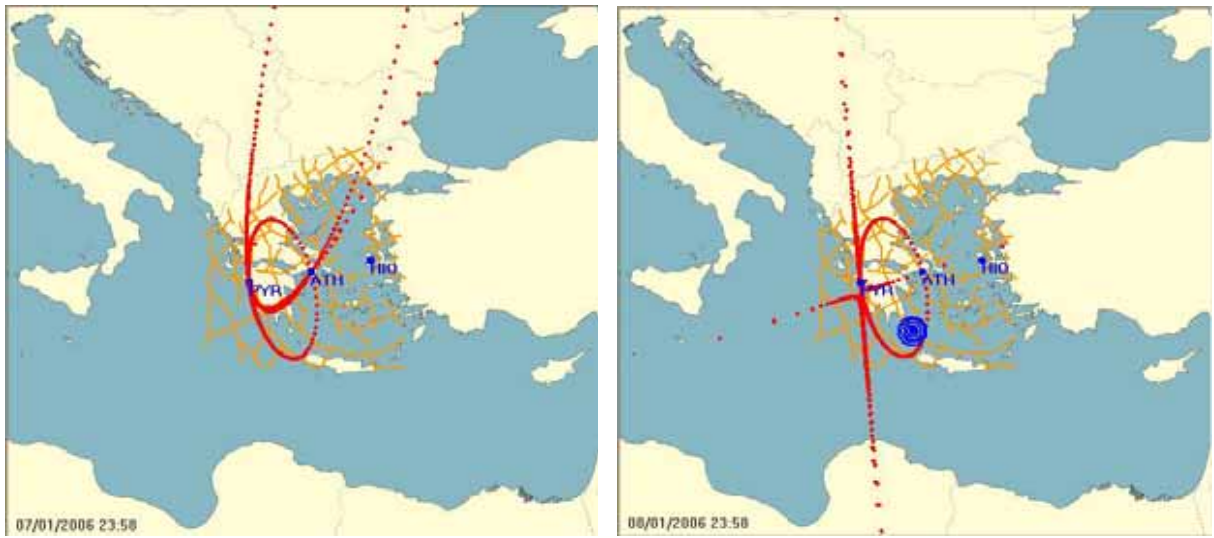

**Fig. 22. Phase maps determined for 20060107 (left) and 20060108 (right). Blue concentric circles indicate the epicenter of the East Kythira EQ.**

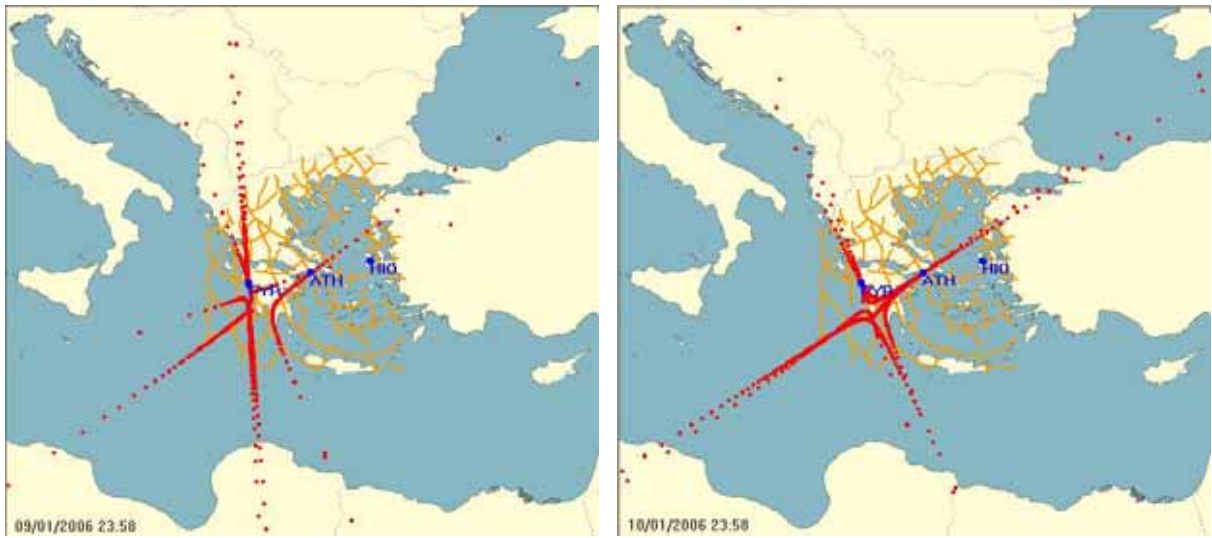

**Fig. 23. Phase maps determined for 20060109 (left) and 20060110 (right).**

In this case too, ellipses were generated for 2 days before the EQ occurrence time, which ellipses vanished right after it.
Finally, the time of occurrence of both EQs is compared to the time of occurrence of the M1 tidal component and the diurnal variation. That comparison is shown in the following figures (24, 25).

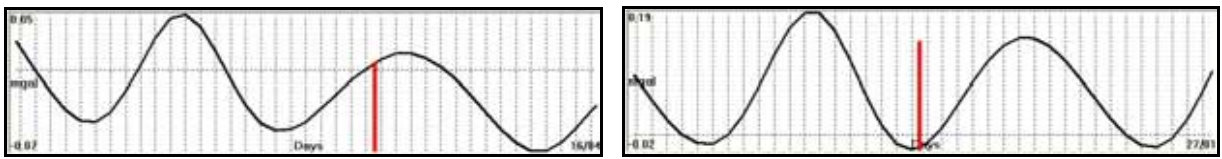

**Fig. 24. M1 tidal variation for 20110401 (left) and 20060108 (right). The red bar indicates the EQ occurrence time.**

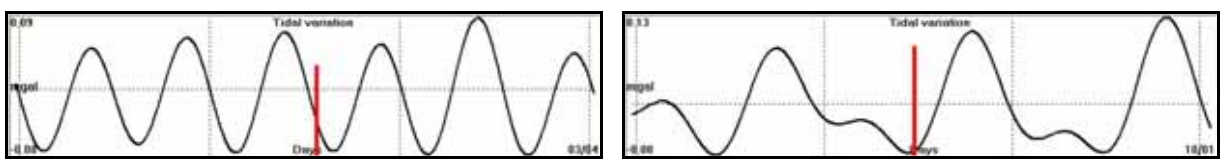

**Fig. 25. Diurnal tidal variation for 20110401 (left) and 20060108 (right). The red bar indicates the EQ occurrence time.**

The fact that two quite large EQs of the Greek territory, behaved in a very similar way, concerning some physical processes, validates, indirectly, the adopted physical models: for the tidally triggered lithospheric oscillation, the generation of the preseismic electric signals and most off all, the fact that the observed generated preseismic electric field in the wider Greek area gives rise to the "strange attractor like" preseismic electric precursors, short before the EQ occurrence (of the order of a couple of days), that is very important for the utilization of a short-term earthquake prediction in terms of occurrence time.

A detailed presentation of the East Kythira EQ concerning its time of occurrence, epicenter location, magnitude determination and observed acceleration deformation prior to the EQ occurrence time have been presented in the monograph "Short term earthquake prediction" and in :www.earthquakeprediction.gr

## 4. References.